\documentclass[cits]{PoS}
\usepackage{xspace}
\newcommand{\Al}{$^{26}$Al\xspace}

\let\jnl=\rmfamily
\def\refe@jnl#1{{\jnl#1}}%
\newcommand\aj{\refe@jnl{AJ}}%
\newcommand\actaa{\refe@jnl{Acta Astron.}}%
\newcommand\araa{\refe@jnl{ARA\&A}}%
\newcommand\apj{\refe@jnl{ApJ}}%
\newcommand\apjl{\refe@jnl{ApJ}}%
\newcommand\apjs{\refe@jnl{ApJS}}%
\newcommand\ao{\refe@jnl{Appl.~Opt.}}%
\newcommand\apss{\refe@jnl{Ap\&SS}}%
\newcommand\aap{\refe@jnl{A\&A}}%
\newcommand\aapr{\refe@jnl{A\&A~Rev.}}%
\newcommand\aaps{\refe@jnl{A\&AS}}%
\newcommand\azh{\refe@jnl{AZh}}%
\newcommand\memras{\refe@jnl{MmRAS}}%
\newcommand\mnras{\refe@jnl{MNRAS}}%
\newcommand\na{\refe@jnl{New A}}%
\newcommand\nar{\refe@jnl{New A Rev.}}%
\newcommand\pra{\refe@jnl{Phys.~Rev.~A}}%
\newcommand\prb{\refe@jnl{Phys.~Rev.~B}}%
\newcommand\prc{\refe@jnl{Phys.~Rev.~C}}%
\newcommand\prd{\refe@jnl{Phys.~Rev.~D}}%
\newcommand\pre{\refe@jnl{Phys.~Rev.~E}}%
\newcommand\prl{\refe@jnl{Phys.~Rev.~Lett.}}%
\newcommand\pasa{\refe@jnl{PASA}}%
\newcommand\pasp{\refe@jnl{PASP}}%
\newcommand\pasj{\refe@jnl{PASJ}}%
\newcommand\skytel{\refe@jnl{S\&T}}%
\newcommand\solphys{\refe@jnl{Sol.~Phys.}}%
\newcommand\sovast{\refe@jnl{Soviet~Ast.}}%
\newcommand\ssr{\refe@jnl{Space~Sci.~Rev.}}%
\newcommand\nat{\refe@jnl{Nature}}%
\newcommand\iaucirc{\refe@jnl{IAU~Circ.}}%
\newcommand\aplett{\refe@jnl{Astrophys.~Lett.}}%
\newcommand\apspr{\refe@jnl{Astrophys.~Space~Phys.~Res.}}%
\newcommand\nphysa{\refe@jnl{Nucl.~Phys.~A}}%
\newcommand\physrep{\refe@jnl{Phys.~Rep.}}%
\newcommand\procspie{\refe@jnl{Proc.~SPIE}}%

\title{The Extreme Sky - Seven Years of INTEGRAL}

\ShortTitle{Workshop Summary}

\author{\speaker{Roland Diehl}\\
        Max Planck Institut f\"ur extraterrestrische Physik, D-85748 Garching, Germany\\
        E-mail: \email{rod@mpe.mpg.de}}

\abstract{Seven years of successful observations of the sky have been completed within the INTEGRAL mission, in the transition regime between X-rays and $\gamma$-rays from $\sim$10-8000~keV. Initially-agreed mission goals have been pursued, and both high-resolution images of point sources and high-resolution spectra of nuclear lines have been obtained. New discoveries have been made, such as X-ray emission from embedded binaries, hard emission tails from AXPs and $^{60}$Fe radioactivity lines; these stimulated both theoretical and observational studies, and now make INTEGRAL a valuable asset for the astronomical survey of high-energy sources across the sky. This contribution summarizes the situation after seven years of the mission, and concludes the 7-year anniversary workshop {\it The extreme sky} held in Otranto, Italy, in Oct 2009.}

\FullConference{INTEGRAL Workshop\\
		{\it The Extreme Universe} \\
		 October 13-17, 2009\\
		 Otranto, Italy}

\begin{document}

\section{INTEGRAL's Origins and Expectations} 
When the INTEGRAL mission was first discussed among scientists, the NASA Compton Gamma-Ray Observatory (CGRO) had just been launched in April 1991 \cite{1993SciAm.269...68G}, beginning its survey of the $\gamma$-ray sky. Developments had led to a change of original plans for CGRO: the fifth originally-planned instrument, a $\gamma$-ray spectroscopy experiment (GRSE), had been abandoned for cost and complexity reasons\cite{1995ExA.....6..129G}. The Compton Observatory performed its all-sky survey for $\gamma$-ray sources in the 100~keV to few GeV range with degree-sized angular resolution and 10-\% sized spectral resolution during a 9-year mission. Simultaneously, and specifically during 1994, proposals were prepared for a next advance in this field, aiming at imaging and spectral resolution improvements by an order of magnitude for this energy range: a coded-mask imaging instrument, and a Ge-detector based spectrometer; the {\it INTEGRAL} mission was given shape \cite{1993SPIE.1945..112S}. The SIGMA instrument \cite{1984AcAau..11..251R} on the GRANAT mission had shown the usefulness of a coded mask for such purpose \cite{1993A&AS...97....1M}, and that same concept was optimized for INTEGRAL's {\it Imager} \cite{1997ESASP.382..599U,2003A&A...411L.131U}. Ge spectrometers had successfully been applied to observations of $\gamma$-ray lines following the spectacular supernova SN1987A in the Large Magellanic Cloud  \cite{1990ApJ...351L..41T}, which led the way to high-resolution $\gamma$-ray spectroscopy. 
The science community welcomed these European efforts for a high-resolution spectrometer, and the US Nuclear-Astrophysics Explorer mission concept (NAE) \cite{1991AdSpR..11..369M} eventually merged with European ideas to form INTEGRAL's {\it Spectrometer} \cite{1997ESASP.382..591M,1998PhST...77...35V,2003A&A...411L..63V}. 

INTEGRAL's mission goals were stated as {\it `` ... spectral $\gamma$-ray features to be uniquely identified and line profiles to be measured for physical studies of the source region, .... and accurate location and hence identification of the $\gamma$-ray emitting objects with counterparts at other wavelengths" }\cite{2003A&A...411L...1W} through {\it `` ...fine spectroscopy with imaging and accurate positioning'' }\cite{2003A&A...411L...1W}. The objects then listed as astrophysical targets were {\it nucleosynthesis, nova and supernova explosions, the interstellar medium, cosmic-ray interactions and sources, neutron stars, black holes, $\gamma$-ray bursts, active galactic nuclei, } and the {\it cosmic $\gamma$-ray background} \cite{2003A&A...411L...1W}. Expectations were great, the {\it quantum leap} for $\gamma$-ray astronomy after the Compton Observatory survey was announced widely. Seven years of successful operations are behind us, and hopefully many more to come. The instruments are all in fine shape and performance in spite of some minor defects, and the healthy spacecraft still has fuel for many more mission years \cite{2009arXiv0912.0077W}. 

\begin{figure}
\centering
\includegraphics[width=0.7\textwidth]{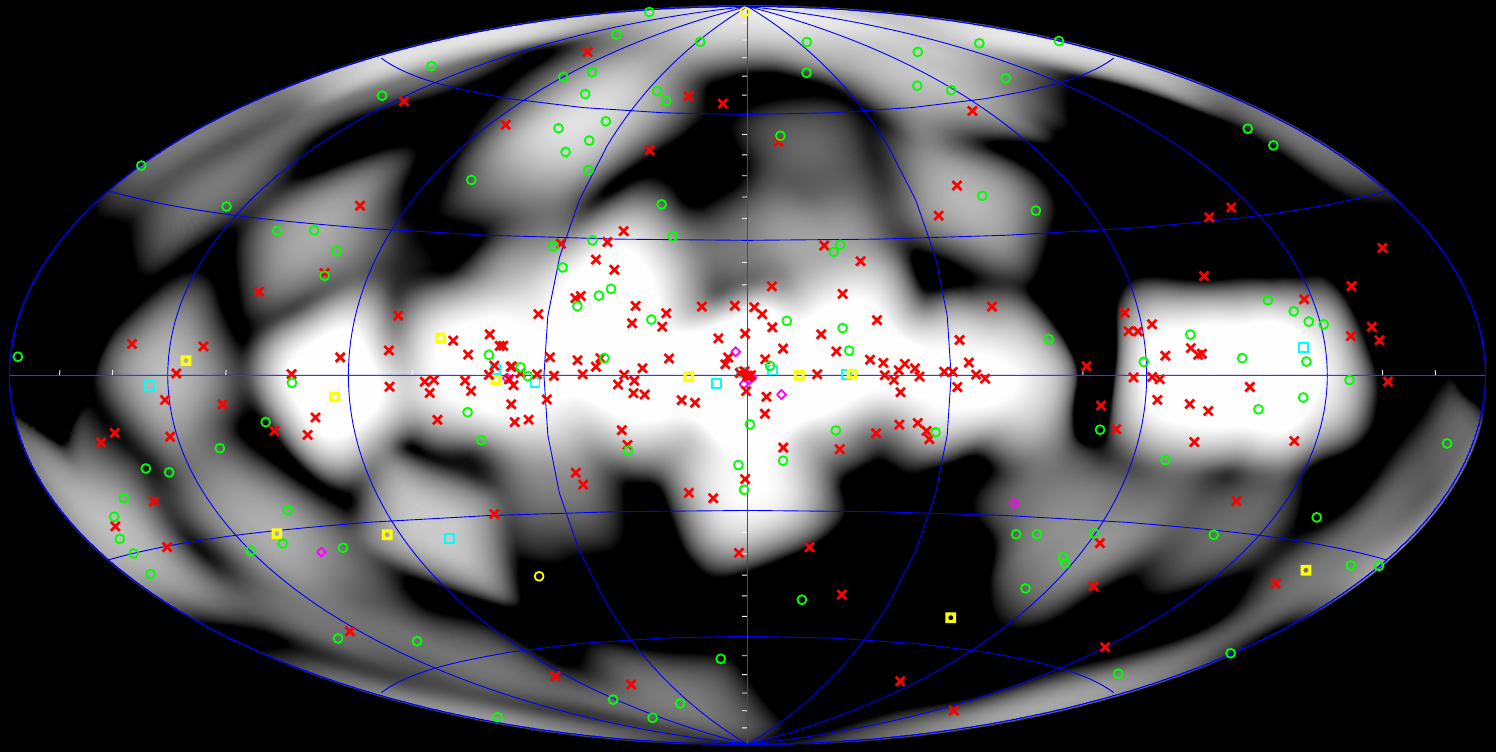}
\caption{The IBIS source catalogue of over 700 sources demonstrates the richness of the sky in $\gamma$-ray sources \cite{2010ApJS..186....1B}. The fine source location capability of 1' or better has helped to identify the majority of these sources (71\%) with known objects, comparing to observations at other wavelengths. }
\label{fig_srccat}
\end{figure}

\section{Scientific Delivery} 

\begin{figure}
\centering
\includegraphics[width=0.28\textwidth]{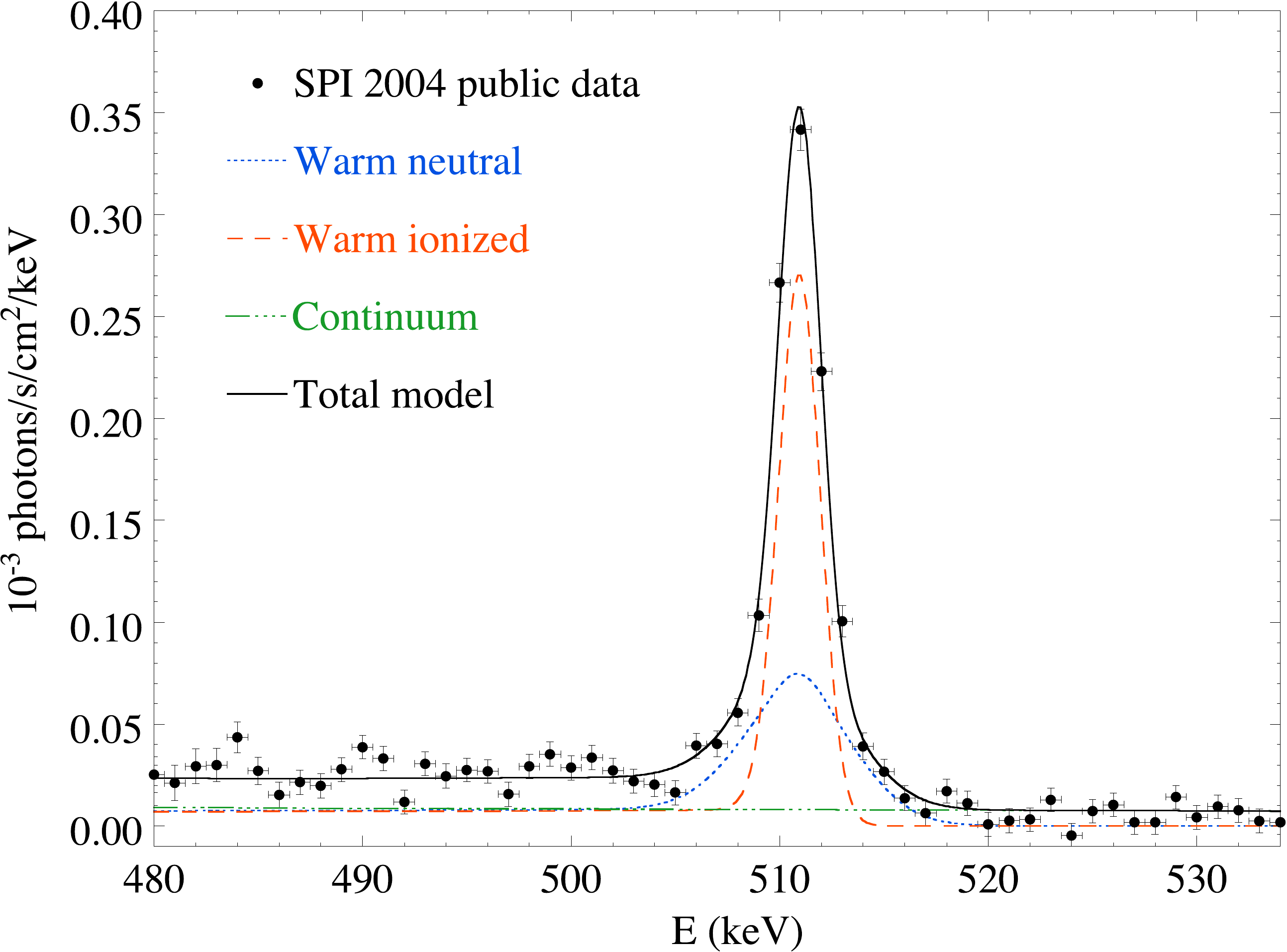}
\includegraphics[width=0.7\textwidth]{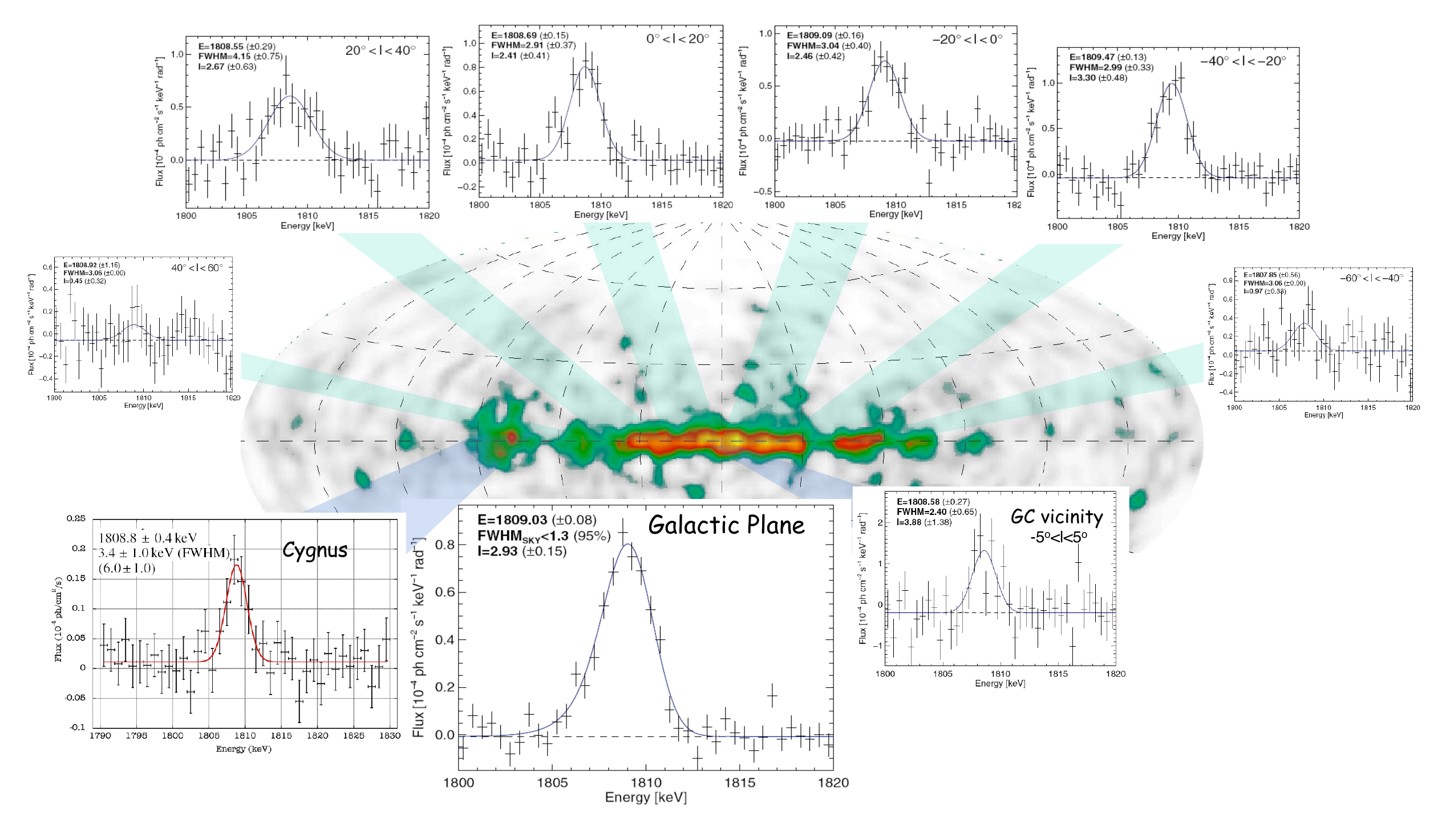}
\caption{The SPI $\gamma$-ray spectrometer resolves astrophysical $\gamma$-ray lines. {\it Left:} Positron annihilation $\gamma$-rays shape the line at 511~keV, as annihilation occurs in cold neutral, warm and partially-ionized, or hot plasma, with expected line widths increasing in that order.  (from  \cite{2006A&A...445..579J}).
{\it Right:} \Al $\gamma$-rays have now been seen from different regions along the plane of the Galaxy, and allow to constrain the recent nucleosynthetic history for different massive-star groups in our Galaxy. (from \cite{2009A&A...496..713W}).}
\label{fig_srccat}
\end{figure}

The first mission years saw impressive statements of scientific delivery --  the variety of papers of this conference reflect latest insights and status summaries. We now have in hand the 4$^{th}$ IBIS catalogue of sources \cite{2010ApJS..186....1B}, which shows a total of almost 700 clearly-significant ($\sim$5$\sigma$) sources in the 17--100~keV range of hard X-rays. With its source location capability of better than 1 arcmin, comparisons to data at other wavelengths helped to identify the nature of the $\gamma$-ray object in 71\% of cases, to tell us about their thermal versus non-thermal emission. 
The $\gamma$-ray line spectroscopy measurements with SPI have obtained clearly-resolved line shape measurements for the strongest $\gamma$-ray line in the sky, the annihilation line of positrons at 511~keV \cite{2004ESASP.552...51J,2005MNRAS.357.1377C}, and \Al line spectra resolved for different source regions along the plane of the Galaxy \cite{2006Natur.439...45D,2009A&A...496..713W}. Line shapes and locations have been interpreted in terms of astrophysical processes of nucleosynthesis sources and interstellar-medium parameters. This clearly demonstrated the {\it delivery} from both of INTEGRAL's main instruments, the Imager and the Spectrometer,  keeping up with the {\it promises} as cited above.  

\section{Some Hard Lessons} 
We had to learn that some things are not as easy as hoped for. SPI's instrumental background turned out to be a big challenge, the hope for efficient rejection through the pulse-shape discrimination system was not fulfilled \cite{2003A&A...411L..91R}. As a result, sensitivity remains below simulated performances. Nuclear de-excitation lines from the interstellar medium ($^{12}$C and $^{16}$O at 4434 and 6129~keV, respectively) could not (yet) be detected. 
Detections of less-bright emissions in annihilation $\gamma$-rays, in \Al, and the discovery of $^{60}$Fe $\gamma$-rays \cite{2007A&A...469.1005W} had to await accumulation of data from five or more mission years. This is beyond the originally-planned duration of even the {\it extended} mission;  but evaluation committees recognized the issues of properly estimating such a dominating and time-variable instrumental background, and generously supported deviations from the original mission plans. The time variations of the spectral response, incurred from detector degradation resulting from cosmic-ray bombardment, and rectified periodically through {\it annealings}, lead to additional efforts in maintaining the spectroscopic precision as required for $\gamma$-ray line shape analyses. 

Gamma-ray lines from supernovae remain a matter of (bad) luck: No sufficiently-nearby ($<$5~Mpc) supernova of type Ia occurred since INTEGRAL's launch, challenging statistical expectations \cite{2008NewAR..52..377I}. The Galactic Cas A supernova remnant's $^{44}$Ti $\gamma$-rays at 68 and 78~keV were seen with the IBIS instrument and thus confirmed earlier measurements \cite{2006ApJ...647L..41R}; but SPI's spectrometer fails to obtain a new measurement of the high-energy line from $^{44}$Ti decay at 1157~keV. Although disappointing at first glance, this can be understood from Doppler-broadening of still-fastly-moving ejecta, causing a broader line to drown in SPI's instrumental background \cite{2009A&A...502..131M}. But the hopes for measuring (and interpreting) the $^{44}$Ti line shape from young supernova remnants could not be fulfilled. Similarly, also INTEGRAL does not detect more of the supernovae that are expected from the Galaxy's current rate of core-collapse supernovae, and only confirms earlier constraints from COMPTEL; no new Cas A-like study objects are found \cite{2006NewAR..50..540R}. 

Also the {\it Imager} instrument encountered difficulties: its spectral response shows a {\it snake-like} nonlinearity which is difficult to control, signal rise-time differences over the large dynamic range of signal amplitudes for the $>$32000 imaging-detector pixels present a major challenge \cite{2005NIMPA.541..323L}. Moreover, seemingly-minor imprecisions in glue applications on the tungsten coded mask holding structure now have been found to limit systematics from artifact features, creeping into images as exposure gets as deep as 70~Ms and beyond. Re- and deepening calibrations are arranged by the instrument and ESA teams, to address these issues within the constraints of the mission. 

\section{Scientific Surprises and New Challenges}   
In spite of these difficulties, INTEGRAL data also presented a number of surprises and new discoveries from the $\gamma$-ray sky:
 
New and unexpected sources were detected in the Galaxy, and identified as sources deeply embedded in surrounding molecular clouds \cite{2003A&A...411L.427W,2006A&A...453..133W,2008A&A...484..783C}. Their intense high-energy emission came as a surprise. Understanding such emission presents a novel challenge. On the other hand, INTEGRAL is able to constrain the contributions from such sources with (non-thermal emission) high-energy tails to the high-energy emission from galaxy clusters, as shown in the Coma cluster \cite{2008ApJ...687..968L,2008A&A...491..209R}.

Binaries with $\gamma$-ray emission resulting from the interaction of a massive-star's intense wind with the orbiting compact companion star were discovered \cite{2005A&A...433L..45K,2009A&A...494L..37H}.
Interestingly, $\gamma$-ray emission was detected from binary sources seen also in TeV high-energy $\gamma$-rays \cite{2005ApJ...630L.157M,2009A&A...494L..37H}, and led to new studies of cosmic-ray acceleration \cite{2005ApJ...629L.109U,2009ApJ...694...12M}. 

Gamma-ray emission from highly-magnetized neutron stars were discovered, and helped to classify the {\it anomalous} X-ray pulsars \cite{2006ApJ...645..556K}. Understanding the magnetosphere as the now-plausible source of high-energy emission remains a challenge, and INTEGRAL data extending well beyond 100~keV in energy are crucial in such study \cite{2008A&A...489..263D}.

The {\it transient} $\gamma$-ray sky also led to surprises \cite{2007A&A...466..595K,2009ApJ...696L..74M}. Magnetar flares such as the INTEGRAL-discovered SGR1806-20 event in 2004 \cite{2005ApJ...624L.105M,2007A&A...476..321E,2007ESASP.622..533G} are one example. 
Other examples arose from $\gamma$-ray burst opportunities \cite{2004AIPC..727..607M}. For GRBs within INTEGRAL's instrument field of views, which occur at a rate of one a month, a few GRBs were seen with {\it harder} spectra than expected from the internal-shock synchrotron model, indicating that the GRB engine is not yet understood. Other GRBs, on the contrary, exhibit spectra which are unusually {\it soft}, and helped to explore the transition range to the phenomena of {\it X-ray flashes}.  Moreover, the anticoincidence system (ACS) of SPI turned out to be useful in GRB studies \cite{2008int..workE...9H}, adding an important GRB monitor with nearly all-sky sensitivity to the interplanetary network, and thus helping GRB locations. 

The inner region of our Galaxy is rich in transient sources, presumably mostly accreting X-ray binaries. INTEGRAL's monitoring program \cite{2007A&A...466..595K} identified and tracked many new such binaries. With now a considerable statistical sample, the spatial and luminosity distributions of these transient sources are being studied \cite{2005A&A...444..821L,2007ASPC..361..388M,2007A&A...461..631N}. It has been recognized that the type of source changes as one steps up into INTEGRAL's $\gamma$-ray domain, with cataclysmic variables and coronally-active stars dominating  the X-ray domain up to several tens of keV, and low-mass X-ray binaries (LMXB) dominating above \cite{2007A&A...471..159R}. The sample of high-mass X-ray binaries could be substantially increased from INTEGRAL's observation of the hard X-ray regime, and now holds nearly hundred objects; prospects for determining their Galactic distribution are realistic now \cite{2005A&A...444..821L}. Within this sample, more sources with unexpected properties are found, such as neutron-star companions with unexpectedly-low rotation periods, or with surprisingly-intense stellar winds \cite{2009MNRAS.398.2152D}. 

Among these HMXBs, a new source class emerged, called {\it superfast X-ray transients} \cite{2007ESASP.622..255N,2006ApJ...646..452S}. In these, the wind interaction with the orbiting compact neutron star apparently sets conditions for intense flaring activity of typically $\sim$hour durations. Constraining the nature of such outbursts with detailed observations of the temporal and spectral changes promises to provide new and unexpected insights into the phenomena and process of accretion of matter onto a compact neutron star, and the nature of the companion star.  

Active galaxies were confirmed to mostly drop sharply in their intensities as one reaches beyond X-ray energies of several keV, as a dusty torus around the active nucleus is believed to absorb radiation. But probing rare active galaxies on the high side of the distribution of absorbing-torus column densities, much fewer such sources were found than had been extrapolated from X-ray observations and theories \cite{2009MNRAS.399..944M}. As a result, the diffuse cosmic X-ray background with its emission maximum at 30--40~keV cannot be inferred by extrapolation of X-ray emission properties, and such extrapolation explains only $\sim$10\% of the peak emission.  

The emission from positron annihilations in interstellar space has been mapped across the sky, and consolidated earlier hints for the bulge region of the Galaxy being by far the brightest emission region on the sky \cite{2005A&A...441..513K,2006A&A...450.1013W,2008Natur.451..159W}. These maps revealed a surprisingly-symmetric bulge emission, and barely were able to detect annihilation emission from the Galaxy's disk, where most of the candidate sources are located (see \cite{2009arXiv0906.1503D} for a  review, and Churazov et al., this volume). 

The \Al line could be seen to vary in centroid position, as would be expected from the Galaxy's large-scale rotation \cite{2006Natur.439...45D}. But such variation of line position is small at tenths of a keV, and it was surprising that INTEGRAL's spectrometer turned out to be able to measure interstellar-medium velocities down to the 100~km~s$^{-1}$ range. This is helped by INTEGRAL's finding that the intrinsic Doppler broadening from ISM kinematics is not as broad as reported before, and instead rather narrow and below instrumental line widths \cite{2006A&A...449.1025D}. Additionally, $^{60}$Fe radioactivity was first clearly measured by INTEGRAL's Spectrometer \cite{2007A&A...469.1005W}.

\section{Prospects for INTEGRAL in its Late Years} 
The INTEGRAL mission is now part of a fleet of astronomy space missions. Coming of age, one may wonder how valuable continued observations with INTEGRAL would be. It is useful to consider the special strengths of this mission for new astrophysical insights.

INTEGRAL's spectrometer SPI features spectral resolution of $\sim$600 (${{E}\over{\delta E}}$). This capability will remain unrivaled and unique for many years to come: Since many astrophysical lines will be kinematically broadened, specifically from the target objects of supernova explosions (see Cas A's experience, as discussed above), the future instrumental developments targeting $\gamma$-ray line studies from supernovae will not need such high resolution. Experiments thus can be optimized towards large collecting areas with less-costly detectors such as CdZn, with adequate spectral resolution for that purpose. 

INTEGRAL targets radiation originating in atomic nuclei from cosmic objects. It is likely that observations in this energy regime have only revealed the tip of the iceberg of cosmic sources, even within our Galaxy. It is worth reminding that most astronomical observations from radio through IR, optical, UV and X-ray bands address emission processes of thermal origins, and spectral information arises from line transitions in the atomic shells. This implies that the state of the atomic shell is part of the emission process, or, stated otherwise, this state must be determined for a proper interpretation of the observed emission. Nuclear emission processes, on the other hand, are often rather independent of the thermodynamic parameters of the emission region, such as in lines originating in radioactive decay, or in high-energy collisions from cosmic rays. This window is uniquely addressed by INTEGRAL and its large field-of-view instruments. With luck, nuclear explosions sufficiently nearby may show the usefulness of such observations of radiation of primarily-nuclear origins.

Polarization of light encodes processes of magnetic-field origins in the cosmic source of radiation. For high-energy radiation, the differential Compton scattering process shows angular patterns which encode polarization, and becomes observable at $\gamma$-ray energies with a sufficiently-large camera. First results obtained with INTEGRAL on a $\gamma$-ray burst and on the Crab pulsar emission are promising \cite{2009A&A...499..465M}. Future observations may further exploit this unique potential.

The INTEGRAL sky survey has emphasized exposure along the plane of the Galaxy. As a result, now deep exposure has been accumulated over regions of the sky which are not as easily accessible by low-orbit missions such as SWIFT. This allows complementarity of sky surveys for active galaxies at high energies (see Krivonos et al., this volume), which can address the issue of which sources are responsible for the cosmic X-ray background in its peak region and above.

Science issues  which appear interesting and hot and can be addressed by future INTEGRAL observations are, for example:
\hfill\break --
What is the Galactic population of high-energy emitting accreting binaries?
\hfill\break --
How does the accretion process occur in close or interacting binaries?
\hfill\break --
What is the nature of high-energy emission in high-field magnetospheres near neutron stars?
\hfill\break --
What does the morphology of positron annihilation emission teach us about positron sources, what about positron escape and interstellar propagation?
\hfill\break --
What are the state and conditions of hot interstellar medium around massive-star groups?
\hfill\break --
How can models of massive-star and supernova structure be aligned with nucleosynthesis of \Al and $^{60}$Fe from those objects?
\hfill\break --
Which active galaxies and their subtypes are responsible for accumulating to the observed diffuse X-ray background?
\hfill\break --
Is emission from $\gamma$-ray bursts, or from $\gamma$-ray pulsars, polarized?
\hfill\break
\noindent
The way an {\it observatory} such as INTEGRAL is managed, some tension arises typically in later mission years from the fact that {\it additional} observations suggested in competitive proposals only add {\it incremental} data to a large and growing database. Therefore, {\it new} insights are not expected, at least from persistent sources of high-energy emission. This appears to argue for an observing program being widely-open for targets of opportunity and tracking of source variabilities, in particular favored by INTEGRAL's large field of view which allow some {\it amalgamation} and {\it serendipidity}.  On the other hand, the unique strengths of INTEGRAL should be utilized to build and consolidate the legacy of results specific to this type of instrumentation, so that astrophysical and deep analysis of these findings (but also future mission proposals to take up and deepen such studies) can be supported with best-achievable precision. 

INTEGRAL is now managed by a {\it Users Group} composed of instrument experts and a broad spectrum of scientists involved in high-energy astrophysics (see also  \cite{2009arXiv0912.0077W}). This group is very effective in discussing the different aspects of observing program alternatives. It turns out a very useful moderator to the diversity of proposed satellite pointings through the annual observing opportunities. Considering the limited gain of such individual and necessarily short (by comparison with existing) exposures, a {\it Key Program} concept has successfully been implemented. Here, two classes of proposal opportunities were installed: {\it Long Key Program} observations set the general program of where INTEGRAL's survey of the $\gamma$-ray sky will be directed to, and a second category of proposals requests {\it data rights} from such prior-selected pointings for analysis of specific astrophysical questions or sources. In a {\it Gedanken-Experiment}, this author also presented a more extreme version of this concept, where exposures are directed by the key program and user group such that synergies are maximized. [One scenario would be to point INTEGRAL at intermediate latitudes along the plane of the Galaxy. This could complete and complement the Galactic-plane survey with the brightest plane regions being in outer field-of-view regions. Simultaneously, exposure would be added to constrain latitudinal extents of diffuse high-energy emission of special interest, such as positron annihilation emission. Active Galaxies could be surveyed at those intermediate latitudes, deepening existing exposures sufficiently for meaningful constraints.] It will remain a challenge to ensure community interest in INTEGRAL observations, especially if the common scheme of regular and frequent {\it observation opportunities} would be deviated from. But a fresh look at the options best-suitable for INTEGRAL could be worthwhile. In any case, INTEGRAL's next seven years could be fruitful to harvest and consolidate the excitements of the extreme universe from its high-energy emission.  

{\bf Acknowledgements}
This review was stimulated from many papers by the active community of high-energy astrophysicists gathered in parts at this nice location at the  beautiful but temporarily unusually-cold coastline of  Puglia. We all appreciated the hosting institutions and generous support of the conference from ESA, Italy's research institutions, and a very energetic organization team. It was a pleasure to celebrate seven years of INTEGRAL in Otranto!

%
\bibliographystyle{plain}
%

\begin{thebibliography}{10}

\bibitem{2010ApJS..186....1B}
A.~J. {Bird}, A.~{Bazzano}, L.~{Bassani}, F.~{Capitanio}, M.~{Fiocchi}, A.~B.
  {Hill}, A.~{Malizia}, V.~A. {McBride}, S.~{Scaringi}, V.~{Sguera}, J.~B.
  {Stephen}, P.~{Ubertini}, A.~J. {Dean}, F.~{Lebrun}, R.~{Terrier},
  M.~{Renaud}, F.~{Mattana}, D.~{G{\"o}tz}, J.~{Rodriguez}, G.~{Belanger},
  R.~{Walter}, and C.~{Winkler}.
\newblock {The Fourth IBIS/ISGRI Soft Gamma-ray Survey Catalog}.
\newblock {\em \apjs}, 186:1--9, January 2010.

\bibitem{2008A&A...484..783C}
S.~{Chaty}, F.~{Rahoui}, C.~{Foellmi}, J.~A. {Tomsick}, J.~{Rodriguez}, and
  R.~{Walter}.
\newblock {Multi-wavelength observations of Galactic hard X-ray sources
  discovered by INTEGRAL. I. The nature of the companion star}.
\newblock {\em \aap}, 484:783--800, June 2008.

\bibitem{2005MNRAS.357.1377C}
E.~{Churazov}, R.~{Sunyaev}, S.~{Sazonov}, M.~{Revnivtsev}, and
  D.~{Varshalovich}.
\newblock {Positron annihilation spectrum from the Galactic Centre region
  observed by SPI/INTEGRAL}.
\newblock {\em \mnras}, 357:1377--1386, March 2005.

\bibitem{2008A&A...489..263D}
P.~R. {den Hartog}, L.~{Kuiper}, and W.~{Hermsen}.
\newblock {Detailed high-energy characteristics of AXP 1RXS J170849-400910.
  Probing the magnetosphere using INTEGRAL, RXTE, and XMM-Newton}.
\newblock {\em \aap}, 489:263--279, October 2008.

\bibitem{2006Natur.439...45D}
R.~{Diehl}, H.~{Halloin}, K.~{Kretschmer}, G.~G. {Lichti},
  V.~{Sch{\"o}nfelder}, A.~W. {Strong}, A.~{von Kienlin}, W.~{Wang}, P.~{Jean},
  J.~{Kn{\"o}dlseder}, J.-P. {Roques}, G.~{Weidenspointner}, S.~{Schanne},
  D.~H. {Hartmann}, C.~{Winkler}, and C.~{Wunderer}.
\newblock {Radioactive $^{26}$Al from massive stars in the Galaxy}.
\newblock {\em \nat}, 439:45--47, January 2006.

\bibitem{2006A&A...449.1025D}
R.~{Diehl}, H.~{Halloin}, K.~{Kretschmer}, A.~W. {Strong}, W.~{Wang},
  P.~{Jean}, G.~G. {Lichti}, J.~{Kn{\"o}dlseder}, J.-P. {Roques}, S.~{Schanne},
  V.~{Sch{\"o}nfelder}, A.~{von Kienlin}, G.~{Weidenspointner}, C.~{Winkler},
  and C.~{Wunderer}.
\newblock {$^{26}$Al in the inner Galaxy. Large-scale spectral characteristics
  derived with SPI/INTEGRAL}.
\newblock {\em \aap}, 449:1025--1031, April 2006.

\bibitem{2009arXiv0906.1503D}
R.~{Diehl} and M.~{Leising}.
\newblock {Gamma-Rays from Positron Annihilation}.
\newblock {\em ArXiv e-prints}, June 2009.

\bibitem{2009MNRAS.398.2152D}
L.~{Ducci}, L.~{Sidoli}, S.~{Mereghetti}, A.~{Paizis}, and P.~{Romano}.
\newblock {The structure of blue supergiant winds and the accretion in
  supergiant high-mass X-ray binaries}.
\newblock {\em \mnras}, 398:2152--2165, October 2009.

\bibitem{2007A&A...476..321E}
P.~{Esposito}, S.~{Mereghetti}, A.~{Tiengo}, S.~{Zane}, R.~{Turolla},
  D.~{G{\"o}tz}, N.~{Rea}, N.~{Kawai}, M.~{Ueno}, G.~L. {Israel}, L.~{Stella},
  and M.~{Feroci}.
\newblock {SGR 1806-20 about two years after the giant flare: Suzaku,
  XMM-Newton and INTEGRAL observations}.
\newblock {\em \aap}, 476:321--330, December 2007.

\bibitem{1995ExA.....6..129G}
N.~{Gehrels}.
\newblock {Hard X-Ray and Gamma-Ray Imaging with Solid State Detectors}.
\newblock {\em Experimental Astronomy}, 6:129--135, December 1995.

\bibitem{1993SciAm.269...68G}
N.~{Gehrels}, C.~E. {Fichtel}, G.~J. {Fishman}, J.~D. {Kurfess}, and
  V.~{Schonfelder}.
\newblock {The Compton Gamma-Ray Observatory}.
\newblock {\em Scientific American}, 269:68--+, December 1993.

\bibitem{2007ESASP.622..533G}
O.~{G{\"o}tz}, S.~{Mereghetti}, K.~{Hurley}, I.~F. {Mirabel}, P.~{Esposito},
  {Tiengo A.}, G.~{Weidenspointner}, and A.~{von Kienlin}.
\newblock {Integral and Magnetars}.
\newblock In {\em ESA Special Publication}, volume 622 of {\em ESA Special
  Publication}, pages 533--+, 2007.

\bibitem{2009A&A...494L..37H}
A.~D. {Hoffmann}, D.~{Klochkov}, A.~{Santangelo}, D.~{Horns}, A.~{Segreto},
  R.~{Staubert}, and G.~{P{\"u}hlhofer}.
\newblock {INTEGRAL observation of hard X-ray variability of the TeV binary LS
  5039/RX J1826.2-1450}.
\newblock {\em \aap}, 494:L37--L40, February 2009.

\bibitem{2008int..workE...9H}
K.~{Hurley}, A.~{Rau}, G.~{Lichti}, and A.~{von Kienlin}.
\newblock {6 years of bursts with the SPI-ACS}.
\newblock In {\em Proceedings of the 7th INTEGRAL Workshop}, 2008.

\bibitem{2008NewAR..52..377I}
J.~{Isern}, E.~{Bravo}, and A.~{Hirschmann}.
\newblock {Detection and interpretation of {$\gamma$}-ray emission from SNIa}.
\newblock {\em New Astronomy Review}, 52:377--380, October 2008.

\bibitem{2006A&A...445..579J}
P.~{Jean}, J.~{Kn{\"o}dlseder}, W.~{Gillard}, N.~{Guessoum}, K.~{Ferri{\`e}re},
  A.~{Marcowith}, V.~{Lonjou}, and J.~P. {Roques}.
\newblock {Spectral analysis of the Galactic positron annihilation emission}.
\newblock {\em \aap}, 445:579--589, January 2006.

\bibitem{2004ESASP.552...51J}
P.~{Jean}, P.~{von Ballmoos}, J.~{Kn{\"o}dlseder}, V.~{Lonjou},
  G.~{Weidenspointer}, G.~K. {Skinner}, M.~{Allain}, E.~{Cisana},
  M.~{Valsesia}, N.~{Guessoum}, R.~{Diehl}, A.~W. {Strong}, M.~{Casse},
  G.~{Vedrenne}, V.~{Sch{\"o}nfelder}, and C.~{Winkler}.
\newblock {Status of the 511 keV Line from the Galactic Centre Region}.
\newblock In V.~{Schoenfelder}, G.~{Lichti}, and C.~{Winkler}, editors, {\em
  5th INTEGRAL Workshop on the INTEGRAL Universe}, volume 552 of {\em ESA
  Special Publication}, pages 51--+, October 2004.

\bibitem{2005A&A...441..513K}
J.~{Kn{\"o}dlseder}, P.~{Jean}, V.~{Lonjou}, G.~{Weidenspointner},
  N.~{Guessoum}, W.~{Gillard}, G.~{Skinner}, P.~{von Ballmoos}, G.~{Vedrenne},
  J.-P. {Roques}, S.~{Schanne}, B.~{Teegarden}, V.~{Sch{\"o}nfelder}, and
  C.~{Winkler}.
\newblock {The all-sky distribution of 511 keV electron-positron annihilation
  emission}.
\newblock {\em \aap}, 441:513--532, October 2005.

\bibitem{2005A&A...433L..45K}
I.~{Kreykenbohm}, N.~{Mowlavi}, N.~{Produit}, S.~{Soldi}, R.~{Walter},
  P.~{Dubath}, P.~{Lubi{\'n}ski}, M.~{T{\"u}rler}, W.~{Coburn},
  A.~{Santangelo}, R.~E. {Rothschild}, and R.~{Staubert}.
\newblock {INTEGRAL observation of V 0332+53 in outburst}.
\newblock {\em \aap}, 433:L45--L48, April 2005.

\bibitem{2006ApJ...645..556K}
L.~{Kuiper}, W.~{Hermsen}, P.~R. {den Hartog}, and W.~{Collmar}.
\newblock {Discovery of Luminous Pulsed Hard X-Ray Emission from Anomalous
  X-Ray Pulsars 1RXS J1708-4009, 4U 0142+61, and 1E 2259+586 by INTEGRAL and
  RXTE}.
\newblock {\em \apj}, 645:556--575, July 2006.

\bibitem{2007A&A...466..595K}
E.~{Kuulkers}, S.~E. {Shaw}, A.~{Paizis}, J.~{Chenevez}, S.~{Brandt},
  {T.~J.-L.} {Courvoisier}, A.~{Domingo}, K.~{Ebisawa}, P.~{Kretschmar}, C.~B.
  {Markwardt}, N.~{Mowlavi}, T.~{Oosterbroek}, A.~{Orr}, D.~{R{\'{\i}}squez},
  C.~{Sanchez-Fernandez}, and R.~{Wijnands}.
\newblock {The INTEGRAL Galactic bulge monitoring program: the first 1.5
  years}.
\newblock {\em \aap}, 466:595--618, May 2007.

\bibitem{2005NIMPA.541..323L}
F.~{Lebrun}, {J.-P.} {Roques}, A.~{Sauvageon}, R.~{Terrier}, P.~{Laurent},
  O.~{Limousin}, F.~{Lugiez}, and A.~{Claret}.
\newblock {INTEGRAL: In flight behavior of ISGRI and SPI}.
\newblock {\em Nuclear Instruments and Methods in Physics Research A},
  541:323--331, April 2005.

\bibitem{2005A&A...444..821L}
A.~{Lutovinov}, M.~{Revnivtsev}, M.~{Gilfanov}, P.~{Shtykovskiy}, S.~{Molkov},
  and R.~{Sunyaev}.
\newblock {INTEGRAL insight into the inner parts of the Galaxy. High mass X-ray
  binaries}.
\newblock {\em \aap}, 444:821--829, December 2005.

\bibitem{2008ApJ...687..968L}
A.~A. {Lutovinov}, A.~{Vikhlinin}, E.~M. {Churazov}, M.~G. {Revnivtsev}, and
  R.~A. {Sunyaev}.
\newblock {X-Ray Observations of the Coma Cluster in a Broad Energy Band with
  the INTEGRAL, RXTE, and ROSAT Observatories}.
\newblock {\em \apj}, 687:968--975, November 2008.

\bibitem{2005ApJ...630L.157M}
A.~{Malizia}, L.~{Bassani}, J.~B. {Stephen}, A.~{Bazzano}, P.~{Ubertini}, A.~J.
  {Bird}, A.~J. {Dean}, V.~{Sguera}, M.~{Renaud}, R.~{Walter}, and
  F.~{Gianotti}.
\newblock {The INTEGRAL/IBIS Source AX J1838.0-0655: A Soft X-Ray-to-TeV
  {$\gamma$}-Ray Broadband Emitter}.
\newblock {\em \apjl}, 630:L157--L160, September 2005.

\bibitem{2009MNRAS.399..944M}
A.~{Malizia}, J.~B. {Stephen}, L.~{Bassani}, A.~J. {Bird}, F.~{Panessa}, and
  P.~{Ubertini}.
\newblock {The fraction of Compton-thick sources in an INTEGRAL complete AGN
  sample}.
\newblock {\em \mnras}, 399:944--951, October 2009.

\bibitem{1993A&AS...97....1M}
P.~{Mandrou}, E.~{Jourdain}, L.~{Bassani}, G.~{Vedrenne}, J.~{Paul}, {J.-P.}
  {Leray}, F.~{Lebrun}, J.~{Ballet}, E.~{Churazov}, M.~{Gilfanov},
  R.~{Sunyaev}, A.~{Bogomolov}, N.~{Khavenson}, N.~{Kuleshova}, I.~{Tserenin},
  and K.~{Sukhanov}.
\newblock {Overview of two-year observations with SIGMA on board GRANAT}.
\newblock {\em \aaps}, 97:1--4, January 1993.

\bibitem{1997ESASP.382..591M}
P.~{Mandrou}, G.~{Vedrenne}, P.~{Jean}, B.~{Kandel}, P.~{Von Ballmoos},
  F.~{Albernhe}, G.~{Lichti}, V.~{Sch{\"o}nfelder}, R.~{Diehl}, R.~{Georgii},
  T.~{Kirchner}, P.~{Durouchoux}, B.~{Cordier}, N.~{Diallo}, F.~{Sanchez},
  B.~{Payne}, P.~{Leleux}, P.~{Caraveo}, B.~{Teegarden}, J.~{Matteson},
  S.~{Slassi-Sennou}, R.~P. {Lin}, G.~{Skinner}, and P.~{Connell}.
\newblock {The INTEGRAL Spectrometer SPI}.
\newblock In {C.~Winkler, T.~J.-L.~Courvoisier, \& P.~Durouchoux}, editor, {\em
  The Transparent Universe}, volume 382 of {\em ESA Special Publication}, pages
  591--+, 1997.

\bibitem{2007ASPC..361..388M}
A.~{Marco}, I.~{Negueruela}, and C.~{Motch}.
\newblock {Blue Stragglers, Be Stars and X-ray Binaries in Open Clusters}.
\newblock In {A.~T.~Okazaki, S.~P.~Owocki, \& S.~Stefl}, editor, {\em Active
  OB-Stars: Laboratories for Stellare and Circumstellar Physics}, volume 361 of
  {\em Astronomical Society of the Pacific Conference Series}, pages 388--+,
  March 2007.

\bibitem{2009A&A...502..131M}
P.~{Martin}, J.~{Kn{\"o}dlseder}, J.~{Vink}, A.~{Decourchelle}, and
  M.~{Renaud}.
\newblock {Constraints on the kinematics of the $^{44}$Ti ejecta of Cassiopeia
  A from INTEGRAL/SPI}.
\newblock {\em \aap}, 502:131--137, July 2009.

\bibitem{2009ApJ...694...12M}
F.~{Mattana}, M.~{Falanga}, D.~{G{\"o}tz}, R.~{Terrier}, P.~{Esposito},
  A.~{Pellizzoni}, A.~{De Luca}, V.~{Marandon}, A.~{Goldwurm}, and P.~A.
  {Caraveo}.
\newblock {The Evolution of the {$\gamma$}- and X-Ray Luminosities of Pulsar
  Wind Nebulae}.
\newblock {\em \apj}, 694:12--17, March 2009.

\bibitem{1991AdSpR..11..369M}
J.~L. {Matteson}.
\newblock {The Nuclear Astrophysics Explorer}.
\newblock {\em Advances in Space Research}, 11:369--378, 1991.

\bibitem{2009A&A...499..465M}
S.~{McGlynn}, S.~{Foley}, B.~{McBreen}, L.~{Hanlon}, S.~{McBreen}, D.~J.
  {Clark}, A.~J. {Dean}, A.~{Martin-Carrillo}, and R.~{O'Connor}.
\newblock {High energy emission and polarisation limits for the INTEGRAL burst
  GRB 061122}.
\newblock {\em \aap}, 499:465--472, May 2009.

\bibitem{2004AIPC..727..607M}
S.~{Mereghetti}.
\newblock {Gamma-Ray Bursts Observed by INTEGRAL}.
\newblock In {E.~Fenimore \& M.~Galassi}, editor, {\em Gamma-Ray Bursts: 30
  Years of Discovery}, volume 727 of {\em American Institute of Physics
  Conference Series}, pages 607--612, September 2004.

\bibitem{2005ApJ...624L.105M}
S.~{Mereghetti}, D.~{G{\"o}tz}, A.~{von Kienlin}, A.~{Rau}, G.~{Lichti},
  G.~{Weidenspointner}, and P.~{Jean}.
\newblock {The First Giant Flare from SGR 1806-20: Observations Using the
  Anticoincidence Shield of the Spectrometer on INTEGRAL}.
\newblock {\em \apjl}, 624:L105--L108, May 2005.

\bibitem{2009ApJ...696L..74M}
S.~{Mereghetti}, D.~{G{\"o}tz}, G.~{Weidenspointner}, A.~{von Kienlin},
  P.~{Esposito}, A.~{Tiengo}, G.~{Vianello}, G.~L. {Israel}, L.~{Stella},
  R.~{Turolla}, N.~{Rea}, and S.~{Zane}.
\newblock {Strong Bursts from the Anomalous X-Ray Pulsar 1E 1547.0-5408
  Observed with the INTEGRAL/SPI Anti-Coincidence Shield}.
\newblock {\em \apjl}, 696:L74--L78, May 2009.

\bibitem{2007A&A...461..631N}
I.~{Negueruela} and M.~P.~E. {Schurch}.
\newblock {A search for counterparts to massive X-ray binaries using
  photometric catalogues}.
\newblock {\em \aap}, 461:631--639, January 2007.

\bibitem{2007ESASP.622..255N}
I.~{Negueruela}, D.~M. {Smith}, J.~M. {Torrej{\'o}n}, and P.~{Reig}.
\newblock {Supergiant Fast X-Ray Transients: A Common Behaviour or a Class of
  Objects?}
\newblock In {\em ESA Special Publication}, volume 622 of {\em ESA Special
  Publication}, pages 255--+, 2007.

\bibitem{2006ApJ...647L..41R}
M.~{Renaud}, J.~{Vink}, A.~{Decourchelle}, F.~{Lebrun}, P.~R. {den Hartog},
  R.~{Terrier}, C.~{Couvreur}, J.~{Kn{\"o}dlseder}, P.~{Martin}, N.~{Prantzos},
  A.~M. {Bykov}, and H.~{Bloemen}.
\newblock {The Signature of $^{44}$Ti in Cassiopeia A Revealed by IBIS/ISGRI on
  INTEGRAL}.
\newblock {\em \apjl}, 647:L41--L44, August 2006.

\bibitem{2006NewAR..50..540R}
M.~{Renaud}, J.~{Vink}, A.~{Decourchelle}, F.~{Lebrun}, R.~{Terrier}, and
  J.~{Ballet}.
\newblock {An INTEGRAL/IBIS view of young Galactic SNRs through the $^{44}$Ti
  gamma-ray lines}.
\newblock {\em New Astronomy Review}, 50:540--543, October 2006.

\bibitem{2008A&A...491..209R}
M.~{Revnivtsev}, A.~{Lutovinov}, E.~{Churazov}, S.~{Sazonov}, M.~{Gilfanov},
  S.~{Grebenev}, and R.~{Sunyaev}.
\newblock {Low-mass X-ray binaries in the bulge of the Milky Way}.
\newblock {\em \aap}, 491:209--217, November 2008.

\bibitem{2007A&A...471..159R}
M.~{Revnivtsev} and S.~{Sazonov}.
\newblock {On the contribution of point sources to the Galactic ridge X-ray
  emission}.
\newblock {\em \aap}, 471:159--164, August 2007.

\bibitem{1984AcAau..11..251R}
G.~{Riviere}, J.~{Paul}, and P.~{Mandrou}.
\newblock {SIGMA: high resolution space observatory project for gamma-ray
  sources.}
\newblock {\em Acta Astronautica}, 11:251--262, 1984.

\bibitem{2003A&A...411L..91R}
J.~P. {Roques}, S.~{Schanne}, A.~{von Kienlin}, J.~{Kn{\"o}dlseder},
  R.~{Briet}, L.~{Bouchet}, P.~{Paul}, S.~{Boggs}, P.~{Caraveo},
  M.~{Cass{\'e}}, B.~{Cordier}, R.~{Diehl}, P.~{Durouchoux}, P.~{Jean},
  P.~{Leleux}, G.~{Lichti}, P.~{Mandrou}, J.~{Matteson}, F.~{Sanchez},
  V.~{Sch{\"o}nfelder}, G.~{Skinner}, A.~{Strong}, B.~{Teegarden},
  G.~{Vedrenne}, P.~{von Ballmoos}, and C.~{Wunderer}.
\newblock {SPI/INTEGRAL in-flight performance}.
\newblock {\em \aap}, 411:L91--L100, November 2003.

\bibitem{2006ApJ...646..452S}
V.~{Sguera}, A.~{Bazzano}, A.~J. {Bird}, A.~J. {Dean}, P.~{Ubertini}, E.~J.
  {Barlow}, L.~{Bassani}, D.~J. {Clark}, A.~B. {Hill}, A.~{Malizia},
  M.~{Molina}, and J.~B. {Stephen}.
\newblock {Unveiling Supergiant Fast X-Ray Transient Sources with INTEGRAL}.
\newblock {\em \apj}, 646:452--463, July 2006.

\bibitem{1993SPIE.1945..112S}
G.~K. {Skinner}, S.~{Bergeson-Willis}, T.~{Courvoisier}, A.~J. {Dean},
  P.~{Durouchoux}, N.~{Eismont}, N.~A. {Gehrels}, J.~E. {Grindlay}, W.~A.
  {Mahoney}, J.~L. {Matteson}, B.~{McBreen}, O.~{Pace}, T.~A. {Prince},
  V.~{Schoenfelder}, R.~{Sunyaev}, B.~{Swanenburg}, B.~J. {Teegarden},
  P.~{Ubertini}, G.~{Vedrenne}, G.~E. {Villa}, S.~{Volonte}, and C.~{Winkler}.
\newblock {INTEGRAL: the next major gamma-ray astronomy mission?}
\newblock In {P.~Y.~Bely \& J.~B.~Breckinridge}, editor, {\em Society of
  Photo-Optical Instrumentation Engineers (SPIE) Conference Series}, volume
  1945 of {\em Presented at the Society of Photo-Optical Instrumentation
  Engineers (SPIE) Conference}, pages 112--123, November 1993.

\bibitem{1990ApJ...351L..41T}
J.~{Tueller}, S.~{Barthelmy}, N.~{Gehrels}, B.~J. {Teegarden}, M.~{Leventhal},
  and C.~J. {MacCallum}.
\newblock {Observations of gamma-ray line profiles from SN 1987A}.
\newblock {\em \apjl}, 351:L41--L44, March 1990.

\bibitem{2005ApJ...629L.109U}
P.~{Ubertini}, L.~{Bassani}, A.~{Malizia}, A.~{Bazzano}, A.~J. {Bird}, A.~J.
  {Dean}, A.~{De Rosa}, F.~{Lebrun}, L.~{Moran}, M.~{Renaud}, J.~B. {Stephen},
  R.~{Terrier}, and R.~{Walter}.
\newblock {INTEGRAL IGR J18135-1751 = HESS J1813-178: A New Cosmic High-Energy
  Accelerator from keV to TeV Energies}.
\newblock {\em \apjl}, 629:L109--L112, August 2005.

\bibitem{1997ESASP.382..599U}
P.~{Ubertini}, G.~{di Cocco}, and F.~{Lebrun}.
\newblock {The IBIS Telescope On Board INTEGRAL}.
\newblock In C.~{Winkler}, T.~J.-L. {Courvoisier}, and P.~{Durouchoux},
  editors, {\em The Transparent Universe}, volume 382 of {\em ESA Special
  Publication}, pages 599--+, 1997.

\bibitem{2003A&A...411L.131U}
P.~{Ubertini}, F.~{Lebrun}, G.~{Di Cocco}, A.~{Bazzano}, A.~J. {Bird},
  K.~{Broenstad}, A.~{Goldwurm}, G.~{La Rosa}, C.~{Labanti}, P.~{Laurent},
  I.~F. {Mirabel}, E.~M. {Quadrini}, B.~{Ramsey}, V.~{Reglero}, L.~{Sabau},
  B.~{Sacco}, R.~{Staubert}, L.~{Vigroux}, M.~C. {Weisskopf}, and A.~A.
  {Zdziarski}.
\newblock {IBIS: The Imager on-board INTEGRAL}.
\newblock {\em \aap}, 411:L131--L139, November 2003.

\bibitem{1998PhST...77...35V}
G.~{Vedrenne}, P.~{Jean}, B.~{Kandel}, F.~{Albernhe}, V.~{Borrel},
  P.~{Mandrou}, J.~P. {Rocques}, P.~{von Ballmoos}, P.~{Durouchoux},
  B.~{Cordier}, N.~{Diallo}, V.~{Sch{\"o}nfelder}, G.~G. {Lichti}, R.~{Diehl},
  M.~{Varendorff}, A.~W. {Strong}, R.~{Georgii}, B.~J. {Teegarden}, J.~{Naya},
  H.~{Seifert}, S.~{Sturner}, J.~{Matteson}, R.~{Lin}, S.~{Slassi},
  F.~{Sanchez}, P.~{Caraeo}, P.~{Leleux}, G.~K. {Skinner}, and P.~{Connell}.
\newblock {The SPI Spectrometer for the INTEGRAL Mission}.
\newblock {\em Physica Scripta Volume T}, 77:35--+, 1998.

\bibitem{2003A&A...411L..63V}
G.~{Vedrenne}, J.-P. {Roques}, V.~{Sch{\"o}nfelder}, P.~{Mandrou}, G.~G.
  {Lichti}, A.~{von Kienlin}, B.~{Cordier}, S.~{Schanne}, J.~{Kn{\"o}dlseder},
  G.~{Skinner}, P.~{Jean}, F.~{Sanchez}, P.~{Caraveo}, B.~{Teegarden}, P.~{von
  Ballmoos}, L.~{Bouchet}, P.~{Paul}, J.~{Matteson}, S.~{Boggs}, C.~{Wunderer},
  P.~{Leleux}, G.~{Weidenspointner}, P.~{Durouchoux}, R.~{Diehl}, A.~{Strong},
  M.~{Cass{\'e}}, M.~A. {Clair}, and Y.~{Andr{\'e}}.
\newblock {SPI: The spectrometer aboard INTEGRAL}.
\newblock {\em \aap}, 411:L63--L70, November 2003.

\bibitem{2003A&A...411L.427W}
R.~{Walter}, J.~{Rodriguez}, L.~{Foschini}, J.~{de Plaa}, S.~{Corbel},
  {T.~J.-L.} {Courvoisier}, P.~R. {den Hartog}, F.~{Lebrun}, A.~N. {Parmar},
  J.~A. {Tomsick}, and P.~{Ubertini}.
\newblock {INTEGRAL discovery of a bright highly obscured galactic X-ray binary
  source IGR J16318-4848}.
\newblock {\em \aap}, 411:L427--L432, November 2003.

\bibitem{2006A&A...453..133W}
R.~{Walter}, J.~{Zurita Heras}, L.~{Bassani}, A.~{Bazzano}, A.~{Bodaghee},
  A.~{Dean}, P.~{Dubath}, A.~N. {Parmar}, M.~{Renaud}, and P.~{Ubertini}.
\newblock {XMM-Newton and INTEGRAL observations of new absorbed supergiant
  high-mass X-ray binaries}.
\newblock {\em \aap}, 453:133--143, July 2006.

\bibitem{2007A&A...469.1005W}
W.~{Wang}, M.~J. {Harris}, R.~{Diehl}, H.~{Halloin}, B.~{Cordier}, A.~W.
  {Strong}, K.~{Kretschmer}, J.~{Kn{\"o}dlseder}, P.~{Jean}, G.~G. {Lichti},
  J.~P. {Roques}, S.~{Schanne}, A.~{von Kienlin}, G.~{Weidenspointner}, and
  C.~{Wunderer}.
\newblock {SPI observations of the diffuse $^{60}$Fe emission in the Galaxy}.
\newblock {\em \aap}, 469:1005--1012, July 2007.

\bibitem{2009A&A...496..713W}
W.~{Wang}, M.~G. {Lang}, R.~{Diehl}, H.~{Halloin}, P.~{Jean},
  J.~{Kn{\"o}dlseder}, K.~{Kretschmer}, P.~{Martin}, J.~P. {Roques}, A.~W.
  {Strong}, C.~{Winkler}, and X.~L. {Zhang}.
\newblock {Spectral and intensity variations of Galactic $^{26}$Al emission}.
\newblock {\em \aap}, 496:713--724, March 2009.

\bibitem{2006A&A...450.1013W}
G.~{Weidenspointner}, C.~R. {Shrader}, J.~{Kn{\"o}dlseder}, P.~{Jean},
  V.~{Lonjou}, N.~{Guessoum}, R.~{Diehl}, W.~{Gillard}, M.~J. {Harris}, G.~K.
  {Skinner}, P.~{von Ballmoos}, G.~{Vedrenne}, J.-P. {Roques}, S.~{Schanne},
  P.~{Sizun}, B.~J. {Teegarden}, V.~{Sch{\"o}nfelder}, and C.~{Winkler}.
\newblock {The sky distribution of positronium annihilation continuum emission
  measured with SPI/INTEGRAL}.
\newblock {\em \aap}, 450:1013--1021, May 2006.

\bibitem{2008Natur.451..159W}
G.~{Weidenspointner}, G.~{Skinner}, P.~{Jean}, J.~{Kn{\"o}dlseder}, P.~{von
  Ballmoos}, G.~{Bignami}, R.~{Diehl}, A.~W. {Strong}, B.~{Cordier},
  S.~{Schanne}, and C.~{Winkler}.
\newblock {An asymmetric distribution of positrons in the Galactic disk
  revealed by {$\gamma$}-rays}.
\newblock {\em \nat}, 451:159--162, January 2008.

\bibitem{2009arXiv0912.0077W}
C.~{Winkler}.
\newblock {INTEGRAL - a status report}.
\newblock {\em ArXiv e-prints}, December 2009.

\bibitem{2003A&A...411L...1W}
C.~{Winkler}, T.~J.-L. {Courvoisier}, G.~{Di Cocco}, N.~{Gehrels},
  A.~{Gim{\'e}nez}, S.~{Grebenev}, W.~{Hermsen}, J.~M. {Mas-Hesse},
  F.~{Lebrun}, N.~{Lund}, G.~G.~C. {Palumbo}, J.~{Paul}, J.-P. {Roques},
  H.~{Schnopper}, V.~{Sch{\"o}nfelder}, R.~{Sunyaev}, B.~{Teegarden},
  P.~{Ubertini}, G.~{Vedrenne}, and A.~J. {Dean}.
\newblock {The INTEGRAL mission}.
\newblock {\em \aap}, 411:L1--L6, November 2003.

\end{thebibliography}

\end{document}